\documentclass[preprint2]{aastex}
\usepackage{amsmath}
\begin{document}
\title{A Review of Optical Sky Brightness and Extinction at Dome C, Antarctica}
\author{S. L. Kenyon and J. W. V. Storey}
\affil{School of Physics, University of New South Wales, Sydney, NSW}
\email{suzanne@phys.unsw.edu.au}
\email{j.storey@unsw.edu.au}
\shorttitle{Sky Brightness at Dome C}
\shortauthors{Kenyon \& Storey}
\affil{Accepted for publication in March 2006 issue of PASP}
\begin{abstract}
The recent discovery of exceptional seeing conditions at Dome C,
Antarctica, raises the possibility of constructing an optical
observatory there with unique capabilities. However, little is known
from an astronomer's perspective about the optical sky brightness and
extinction at Antarctic sites. We review the contributions to sky
brightness at high-latitude sites, and calculate the amount of usable
dark time at Dome C. We also explore the
implications of the limited sky coverage of high-latitude sites and
review optical extinction data from the South Pole. Finally, we examine the
proposal of \citet{Baldry2001} to extend the amount of
usable dark time through the use of polarising filters.
\end{abstract}
\section{Introduction}
Dome C, Antarctica, is one of the most promising new sites for
optical, infrared and sub-millimeter astronomy. Located at
$75^\circ6'$ South, $123^\circ21'$ East and an altitude of 3250 m,
Dome C is the third highest point on the Antarctica Plateau. The
French (IPEV) and Italian (PNRA) Antarctic programs have operated a
summertime scientific base on Dome C since 1995
\citep{Candidi2003}. Construction of a wintertime station was
completed at the beginning of 2005, leading to the first manned winter
season at Dome C. Preliminary site testing has been carried out at
Dome C since 1995 \citep{Valenziano1999, Candidi2003} and systematic
measurements of the summertime \citep{Aristidi2005b} and wintertime
\citep{Storey2005} characteristics of the site began in 2001. The
results, summarised below, indicate very favourable conditions for
astronomy. 

The local topography of Dome C is extremely flat, resulting in a mean
ground-level wind speed of 2.9 ms$^{-1}$ \citep{Aristidi2005}, less
than half that at most other observatories. Wintertime
measurements of the turbulence in the atmosphere above
Dome C (with a MASS and a SODAR) indicate that, above a thin
surface layer, the atmosphere is extremely stable. From above 30
metres, a median seeing of 0.27$\arcsec$ is observed during the
winter, with the seeing below 0.15$\arcsec$ for 25\% of the time
\citep{Lawrence2004}. In comparison, the median seeing at the best
mid-latitude sites is between 0.5 and 1.0$\arcsec$. \citet{Agabi2005}
measure the wintertime seeing from the ground level to be
$1.9\pm0.5\arcsec$ and attribute this to a thin but intense layer of
surface turbulence. From above this layer, which they estimate to be
36 m, the exceptionally good seeing observed by \citet{Lawrence2004}
is confirmed.  In addition, summertime site testing with a DIMM, 
 i.e. while the sun is continuously above the horizon, shows a
 remarkable median seeing of 0.54$\arcsec$ \citep{Aristidi2005b}. 
\begin{deluxetable}{lcccc}
  \tablewidth{469.6 pt}
 \tablecaption{Observable sky \label{tab:sky}}
 \tablehead{\colhead{Site} & \multicolumn{2}{c}{45$^\circ$ zenith limit} & \multicolumn{2}{c}{60$^\circ$ zenith limit}\\ & Dec range & Percentage of sky& Dec range & Percentage of sky}
 \startdata
Equator            &  $45^\circ$ S -- $45^\circ$ N &  70 \%  &  $60^\circ$ S -- $60^\circ$ N & 86 \%\\
Mauna Kea          &  $26^\circ$ S -- $64^\circ$ N &  66 \%  &  $41^\circ$ S -- $79^\circ$ N & 81 \%\\
Dome C             &  $90^\circ$ S -- $30^\circ$ S &  25 \%  &  $90^\circ$ S -- $15^\circ$ S & 37 \%\\
North or South Pole&  $90^\circ$ N/S -- $45^\circ$ N/S &  14 \%  &  $90^\circ$ N/S -- $30^\circ$ N/S & 25 \%
\enddata
\end{deluxetable}

The cloud cover is very low with cloud-free skies observed for at
least 74\% of the time \citep{Ashley2005}. In addition, the site is
extremely cold and the atmosphere has very low precipitable water
vapour in comparison to other sites \citep{Valenziano1999}, leading to
exceptionally low sky backgrounds in the infrared \citep{Walden2005}
and the sub-millimeter \citep{Calisse2004}. Substantial improvements
in atmospheric transmission are predicted at these wavelengths and a
number of new spectral windows are opened up that are inaccessible at
non-Antarctic sites \citep{Lawrence2004a}. 

Despite these attractions, the high latitude of Dome C means that the
sun spends a relatively small amount of time far below the
horizon. This implies longer periods of astronomical twilight and less
\emph{optical} dark time than other sites, especially those close to
the equator. Thus, although the advantages
offered by Antarctica at infrared wavelengths and beyond (where sky
brightness is unaffected by twilight, moonlight and aurorae) are well
established (e.g. \citealt{Storey2003}), it remains to be seen whether the amount of usable
dark time at Dome C is sufficient to allow useful advantage to be taken
of the seeing conditions in the optical. 

In addition to the long twilight, high latitude sites suffer from 
reduced sky coverage. The fraction $f$ of the 
total $4\pi$ steradians of the heavens that can be seen from a site of 
given latitude over a 24 hour period is
\begin{equation}
f=1/2(\sin N - \sin S)
\end{equation}
where $N$ is the northernmost declination observable and $S$ is the 
southernmost. We consider two cases: a zenith limit of 45 degrees (1.4 
airmasses), and a zenith limit of 60 degrees (2 airmasses). As seen in 
Table \ref{tab:sky}, Mauna Kea has access to 81\% of the sky at 2
airmasses or less, while at Dome C only 37\% of the sky is similarly
available. 
\placetable{tab:sky}

In the case of a high-latitude southern site such as Dome C, this 
restricted sky coverage is mitigated somewhat by the accessibility of 
several key sources such as the Large (dec = $69^\circ$ S) and Small (dec =
$73^\circ$ S ) Magellanic Clouds and the 
Galactic Centre (dec $=28^\circ$ S). Such southern sources are of
course favourably observed from Dome C. For example, although the
Galactic Centre reaches comparable maximum elevations of 44$^\circ$ at
Dome C and 41$^\circ$ at Mauna Kea, it is above
30$^\circ$ elevation for 1300 dark hours per year at Dome C, but only
660 dark hours per year at Mauna Kea. The advantages of a
high-latitude southern site for the continuous monitoring of southern
objects have already been discussed by a number of authors; for
example, \citet{Deeg2005}. 

In Section \ref{sec:brightness}, we assess the amount of \emph{usable} dark time
at Dome C and examine the various contributions to optical sky brightness. The
causes of atmospheric extinction are discussed in Section
\ref{sec:atmos}. \citet{Baldry2001} have suggested using a polarising 
filter to reduce twilight contributions to sky background. In Section
\ref{sec:pol} we examine the polarisation of scattered sunlight over
the course of twilight to evaluate the feasibility of this idea. 

\section{Optical sky brightness}\label{sec:brightness}
\placetable{tab:contributions}
\begin{deluxetable}{lll}
  \tablecaption{Contributions to the light of the night sky \label{tab:contributions}}
  \tablehead{\colhead{Source} & \colhead{Function of} & \colhead{Physical Origin}}
  \startdata
Scattered sunlight & Ecliptic coordinates, season,   &  Sunlight scattering from molecules\\
                   & location, aerosols                & and  particles in the upper atmosphere\\ 
Moonlight          & Lunar phase, position           & Sunlight reflected from the lunar\\
                   & of the moon, aerosols           & surface, then scattered in the atmosphere\\
Aurora             & Magnetic latitude, season,      & Excitation of upper atmosphere atoms  \\
                   & magnetic activity, solar activity,& and molecules by energetic particles \\
Airglow            & Zenith angle, local time,        & Chemiluminescence of upper \\
                   & latitude, season, solar activity,  & atmosphere atoms and molecules \\
                   & altitude, geomagnetic latitude            & \\
Zodiacal light     & Ecliptic coordinates              & Sunlight scattered by interplanetary dust \\
Integrated starlight   & Galactic coordinates          & Unresolved stars in the Milky Way\\
Diffuse galactic light & Galactic coordinates          & Scattering of starlight by interstellar dust\\
Integrated cosmic light& Galactic coordinates,   & The universe\\
                       & cosmological red shift&  \\
Light pollution  & Proximity to civilisation           & Artificial lighting\\
  \enddata
\tablecomments{From \cite{Roach}}
\end{deluxetable}

The night sky is never completely dark at any site. The darkest
optical skies on Earth are of the order of 22.0 -- 21.1 V mag
arcsec$^{-2}$ at zenith \citep{Leinert1998}. See \citet{Leinert1998}
for a comprehensive discussion of diffuse night sky brightness,
\citet{Patat2003} for an in depth survey of \emph{UBVRI} night sky
brightness at ESO-Paranal, and \citet{Benn1998} for a review of sky
brightness measurements at La Palma. 

On a moonless night, long after the sun has set, the sky is brightened
by aurorae, airglow, zodiacal light, integrated starlight, diffuse
galactic light, extra-galactic light and artificial sources (see Table
\ref{tab:contributions}). As first noted by
\citet{Rayleigh1928} the sky brightness at a particular site varies
with solar activity; for example, \citet{Walker1988} records
a change of at least 1 mag arcsec$^{-2}$ in sky brightness with solar activity.
Atmospheric scattering of the
flux from each of these sources adds significantly to the sky
brightness; for example, the  contribution from zodiacal light can be
increased by more than  15\% by scattering and that from integrated
starlight by 10 -- 30\% \citep{Leinert1998}. During twilight, the
scattered sunlight usually makes the dominant contribution to the
overall sky brightness, while the
flux from direct and scattered moonlight adds a further strong
contribution when the moon is above the horizon. The 
best dark conditions occur at a site with minimal atmospheric scattering. 

In this section we discuss each source of sky brightness at Dome C
and provide a comparison to Mauna Kea, Hawaii, which we select because
it is the closest major observatory to the equator.

\subsection{Scattered sunlight and usable dark time}
Sunlight is the strongest contributor to the brightness of the
sky. During the day the surface of the Earth is illuminated by
\emph{direct} sunlight and sunlight \emph{scattered} by atmospheric
molecules and particles. After sunset, the surface of the Earth is only
illuminated by scattered sunlight; the direct component illuminates
the atmosphere above the level of the Earth's shadow. Over the course
of twilight the scattered sunlight contribution to sky brightness
decreases, as higher and less dense levels of the atmosphere are
illuminated \citep{Rozenberg}. In
the absence of moonlight and artificial sources, scattered sunlight
completely dominates the sky brightness until 
the sun reaches a zenith angle of about 98 degrees
\citep{Pavlov1995}. The relative contribution 
of scattered sunlight to total sky brightness then 
decreases sharply to a negligible level, and nighttime starts. In the
V-band this occurs when the sun is depressed $17$ -- $19^\circ$ below the
horizon; if there is a high amount of scattering in the atmosphere this can
be extended to depressions of 20 -- 23$^\circ$
\citep{Rozenberg}. Astronomical nighttime is usually considered to 
begin when the sun is more than $18^\circ$ below the horizon.

For solar zenith angles greater than about $98^\circ$, multiple
scattering accounts for essentially all of the sunlight contribution
to sky brightness; this contribution is affected by the aerosol
content in the atmosphere \citep{Pavlov1995, Ugolnikov2004}. Dome C has
an exceptionally clear atmosphere with no dust, haze, smog or sand
aerosols (see Section \ref{sec:atmos} for a discussion on scattering
at Dome C) and therefore it is expected that the intensity of multiply
scattered light will be at a minimum possible level. Paradoxically, the
lack of aerosols will, however, decrease the attenuation of sunlight
that grazes the earth, with the result that at mid-twilight the
intensity just above the earth's shadow could be brighter at some
wavelengths than at sites with less clear skies. This has the effect
of lowering the effective altitude of the so-called ``twilight layer''
\citep{Oug1999, Ougolnikov2003}, creating an opportunity for enhanced scattering
by the denser gas at those lower altitudes.  A further effect that
could brighten the Dome C sky is the high albedo of the snow that
covers the Antarctic plateau, which will increase the sky brightness
by a small amount at sunset.  However this effect decreases with
increasing solar zenith angles and is negligible once the sun has set
\citep{Anderson1990}. During periods when the moon is above the
horizon, the high albedo of the snow may also increase the sky
brightness beyond that normally experienced during grey time.

From the above discussion, it is clear that the twilight sky brightness 
at Dome C and the solar depression angle at which nighttime commences 
cannot be calculated in a straightforward manner. Both parameters can 
only be quantified by direct measurement or by detailed modelling using 
accurate atmospheric aerosol profiles.
\begin{figure}[h]
\plotone{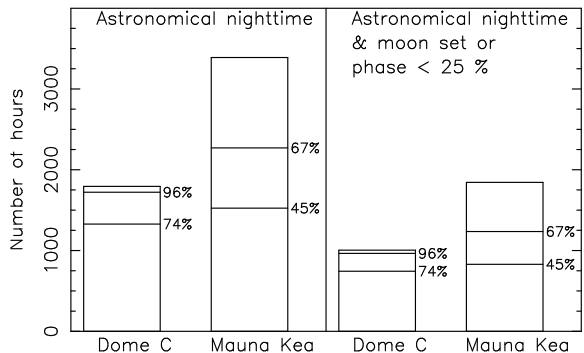}
\caption{Comparison of the available dark time and cloud statistics
  (see text) at Dome C and Mauna Kea. The left panel shows the number
  of hours of formal astronomical nighttime (i.e. when the sun is
  further than $18^\circ$ below the horizon). The right panel shows
  the number of hours of astronomical nighttime when the moon is below
  the horizon or less than one quarter phase. The percentage bars indicate
  the usable dark time once cloud cover is taken into account. 
\label{fig:time}}  
\end{figure}

Using the formal definition above, Dome C has less astronomical nighttime
than lower latitude sites, with long periods of twilight. From a
simple geometric calculation, Dome C has about 50\% of the
astronomical nighttime of Mauna Kea, as shown in Figure
\ref{fig:time}. The cloudiness of each site also needs to be taken
into consideration when looking at the amount of dark time available
for optical observations. On the basis of 2001 data, \citet{Ashley2005} report the skies at Dome
C to be cloud-less for 74\% of the time, with the remaining 26\%
having more than 1/8 cloud cover. While this is a tentative conclusion
based on less than a full year of data from Dome C, new independent
measurements in 2005 \citep{Aristidi2005c} suggest that the percentage
of cloud-free skies at Dome C may be as high as 96\%. In comparision
\citetext{\citealt{Ortolani2003} and references therein} report 45\%
photometric nights (no clouds) and 67\% spectroscopic nights (1/4 -- 1/10 cloud
cover) at Mauna Kea. These percentages are also displayed in Figure
\ref{fig:time}.  

If the more optimistic figure of \citet{Aristidi2005c} is confirmed,
Dome C will be seen to have a comparable number of usable,
astronomically dark hours to Mauna Kea.
\placefigure{fig:time}

\subsection{Moonlight}
\begin{figure}[p]
  \plotone{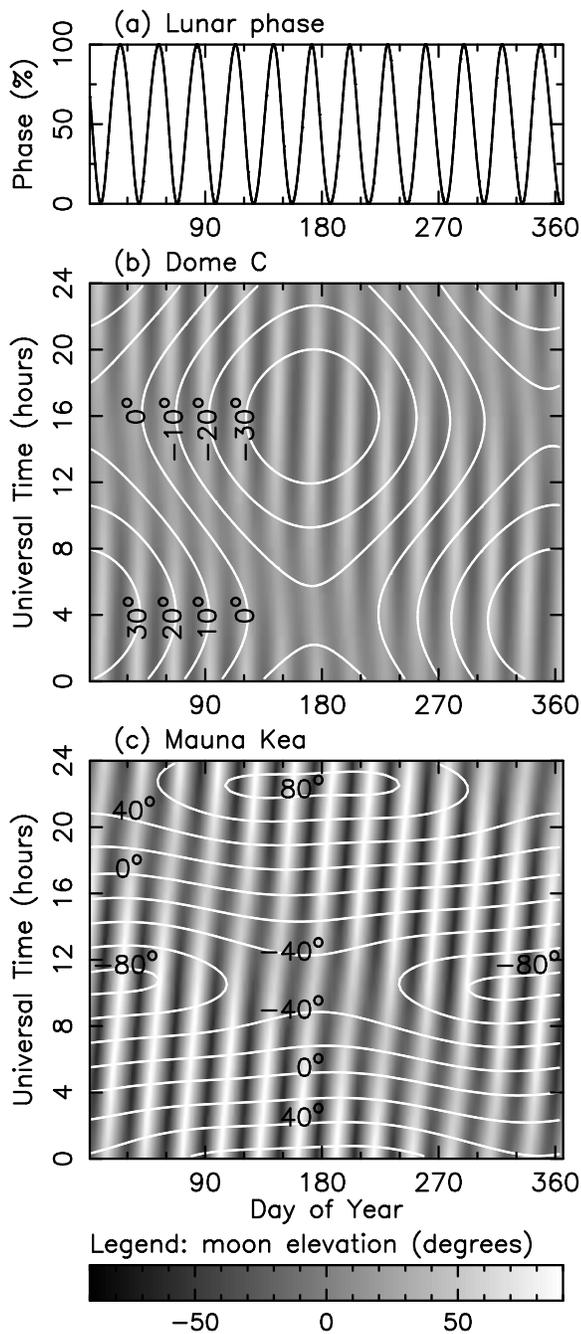}
  \caption{Lunar phase (a), and solar and lunar elevation angles
  at Dome C (b) and Mauna Kea (c) over the course of one year
  (2005). In plots (b) and (c) the contours show the solar elevation
  angles and the grey scale shows the lunar elevation angles according
  to the scale below.\label{fig:sunelevation} } 
\end{figure}

\begin{figure}[h]
\plotone{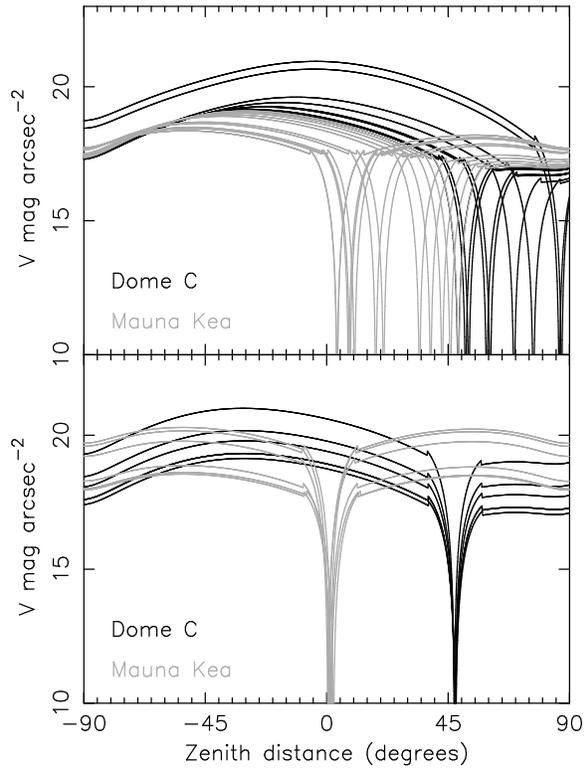}
\caption{Top panel: Scattered moonlight brightness as a function of zenith angle
  when the \emph{full} moon
  reaches its highest altitude in each month (in 2005). Bottom panel: Scattered moonlight
  contribution for the highest altitude of the moon in each month,
  regardless of phase. Each plot shows the moonlight contribution for a cross
  section of the sky that passes through the moon and the zenith, for Dome C (black)
  and Mauna Kea (grey), and each curve represents a different
  month. \label{fig:moonlight}} 
\end{figure}

The sky brightness contribution caused by the Moon depends on its
position and phase. Moonlight illuminates the surface of the Earth
both directly and by scattering, in a similar fashion to sunlight. Figure
\ref{fig:sunelevation} shows the elevation of the sun (contour lines)
and moon (gray scale shading), together with the phase of the moon,
for Dome C and Mauna Kea over the course of one year (2005). The darkest
skies are when the sun is far below the horizon and
either the moon is below the horizon (dark shading) or the phase of the moon is
close to new. The 18.6 year lunar nodal cycle changes the declination range of
the moon; this ranges between $\pm 29^\circ$ and $\pm
18^\circ$. At Dome C the moon reaches a maximum elevation between
$\sim 33^\circ$ and $\sim 43^\circ$ depending on this cycle; at Mauna
Kea the moon can pass through the zenith, independent of nodal cycle. 
\placefigure{fig:sunelevation}
\placefigure{fig:moonlight}

To quantify the effect of the different range of moon elevation at
Dome C and Mauna Kea, we calculated the moonlight brightness using the
model of \citet{Krisciunas1991}. The lunar phase (from full moon to
new moon and back) cycles over one 
month; for each month in 2005 we calculated the sky brightness when (1) the
\emph{full} moon was at the highest elevation for that month and (2)
the moon was at the highest altitude, regardless of phase. A cross
section of the sky, cutting through the position of the moon and the
zenith at those times, is plotted in Figure \ref{fig:moonlight}. In
the calculation we set the extinction coefficient for both sites to
the median value for Mauna Kea (0.12 mag/airmass at 550 nm).

The full-moon contribution at Dome C is 
less than that at Mauna Kea by several magnitudes at zenith,
with little difference at the horizon. For the second case, excluding
the sky close to the moon, the contribution is once again less at Dome
C than at Mauna Kea. As at all sites, this contribution reaches a
minimum between $60^\circ$ and $90^\circ$ away from the moon
\citep{Patat2004}. This advantage to Dome C is reduced to some extent
by the fact that the fullness of the moon and its maximum elevation
are highly correlated at Dome C (as seen in Figure
\ref{fig:sunelevation}) but only weakly so at Mauna Kea. Skies that
would otherwise be dark are brightened at zenith by moonlight by a
median value of 1.7 V mag arcsec$^{-2}$ at Dome C and 2.1 V mag
arcsec$^{-2}$ at Mauna Kea, averaged over the epoch 2005 -- 2015.

\subsection{Aurorae}
\placetable{tab:aurora}
\begin{deluxetable}{ccccccc}
  \tablewidth{0 pt}
  \tablecolumns{6}
  \tablecaption{Position of aurorae at Dome C \label{tab:aurora}}
 \tablehead{&& \multicolumn{2}{c}{Aurora altitude: 100 km} & &
  \multicolumn{2}{c}{Aurora altitude: 250 km}\\ Angular separation  &&
  $D$ (km) & $E$ && $D$ (km)  & $E$}
  \startdata
$10^\circ$ &&	1125	&	\phs$0^\circ$ &&	1161	&	\phs$7^\circ$ \\	
$11^\circ$ &&	1236	&	$-1^\circ$ &&	1271	&	\phs$6^\circ$ \\	
$12^\circ$ &&	1347	&	$-2^\circ$ &&	1382	&	\phs$4^\circ$ \\	
$13^\circ$ &&	1459	&	$-3^\circ$ &&	1493	&	\phs$3^\circ$ \\	
$14^\circ$ &&	1570	&	$-3^\circ$ &&	1604	&	\phs$2^\circ$ \\	
$15^\circ$ &&	1681	&	$-4^\circ$ &&	1716	&	\phs$1^\circ$ \\	
$16^\circ$ &&	1792	&	$-5^\circ$ &&	1827	&	\phs$0^\circ$ \\	
$17^\circ$ &&	1903	&	$-6^\circ$ &&	1938	&	$-1^\circ$ 	
\enddata
\tablecomments{The first column is the angular separation between the aurora and
  Dome C. From this we derive the distance, $D$, from the Dome C to the
  aurora and the elevation angle, $E$, of the aurora at Dome C.}
\end{deluxetable}

\begin{figure}
\plotone{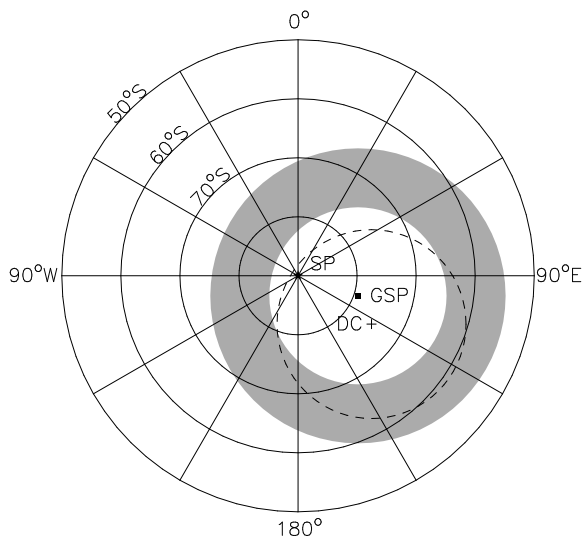}
\caption{Schematic of the typical southern auroral oval, showing the
  geographic South Pole (SP), Dome C (DC) and the geomagnetic south
  pole (GSP). The auroral oval, shown in grey shading, is typically located 15
  -- 25$^\circ$ from the geomagnetic south pole (GSP). Aurorae at 250
  km altitude will be above the horizon at Dome C only if they lie within
  the dotted line.\label{fig:aurora}} 
\end{figure}

Aurorae are the spectacular lights seen dancing across the skies in
(mainly) polar regions. As highly energetic particles from the sun
travel into the atmosphere they collide with upper atmosphere 
atoms and molecules, exciting them to higher energy levels. The
excited atoms and molecules then decay radiatively. Aurorae are
generally (though not exclusively) confined to an annular region 15 --
25 degrees from the \emph{geomagnetic} 
poles (see Figure \ref{fig:aurora}) and have a strong dependence on
the 11 year solar sunspot cycle. The geomagnetic poles are the two
positions where the Earth's theoretical magnetic dipole intersects the Earth's surface. The
positions of the geomagnetic poles move around, and in January 2005
the geomagnetic south pole was at a geographic position of 79.3$^\circ$
S, 108.5$^\circ$ E \citep{geo}. The size and position of the auroral
oval changes with solar activity; see \citet{ovation} for plots of the
size and position of the auroral oval from December 1983 to the
present. Aurorae typically occur 100 -- 250 km above the ground but
can occur at altitudes anywhere between 80 and 1000 km. In 
the V-band the strongest auroral emission is from neutral oxygen at
557.7 nm. Aurorae can vary rapidly in intensity and position across
the sky.
\placefigure{fig:aurora}
Dome C lies about 10$^\circ$ away
from the inner edge of the typical auroral oval, so little auroral activity is
expected. In comparison, South Pole is located very close to the inner
edge of the auroral oval and experiences considerable auroral activity. The
position of Dome C relative to the auroral oval 
means that most auroral activity will be close to the horizon as well
as low in intensity. Figure \ref{fig:aurora} shows the geographic
positions of Dome C, South Pole (geographic), the typical auroral oval
and the geomagnetic south pole. Using simple geometry we 
calculated the sky position at Dome C for aurorae occurring at 100 and
250 km altitude in the auroral oval, as a function of angular
separation between Dome C and the aurora. Aurorae
at 100 km altitude will usually be below the horizon at Dome C; 
at 250 km altitude this occurs at 16$^\circ$ separation. The dotted circle
in Figure \ref{fig:aurora} shows this range. The results of the
calculations are shown in Table \ref{tab:aurora}. The first column is
the angular separation between the aurora and Dome C, $D$ is
the distance in kilometres from Dome C to the
aurora, and $E$ is the elevation angle of the aurora at Dome C. The elevation
angle of typical aurorae seen from Dome C will be between the horizon and
7$^\circ$ elevation, and they will be between 1160 and 2000 km away. Sun-aligned 
quiet arcs will not decrease with increasing magnetic latitude in the
same fashion as normal aurorae, however they are significantly less
intense than the aurorae occurring within the auroral oval (Gary Burns,
2005, private communication).

\citet{Dempsey2005} used satellite measurements of the electron flux
above Dome C to calculate the expected intensity of aurorae. They
found that in the V band the intensity of the auroral contribution to
sky background was less than 22.7 mag arcsec$^{-2}$ for 50\% of the
wintertime during a solar maximum year and below 23.5 mag arcsec$^{-2}$
during solar minimum. 

Aurorae are therefore expected to have a minor impact on optical
astronomy at Dome C, even without the use of narrow-band filters to
remove the brightest emission lines. The contribution to sky
brightness at Mauna Kea by aurorae is of course negligible.

\subsection{Zodiacal light}
Zodiacal light is caused by sunlight scattering from the diffuse cloud
of interplanetary dust that lies largely in the plane of the solar
system. Zodiacal light is strongest near the sun and is generally seen as a
cone of light with its base at the horizon, decreasing in intensity towards
the zenith, with another maximum sometimes seen at the anti-solar point. The
intensity is dependent on wavelength, observer position and sky
position. Zodiacal light is polarised, reaching a maximum polarisation
of about 20\% \citep{Leinert1998}.

\begin{figure}[h]
  \plotone{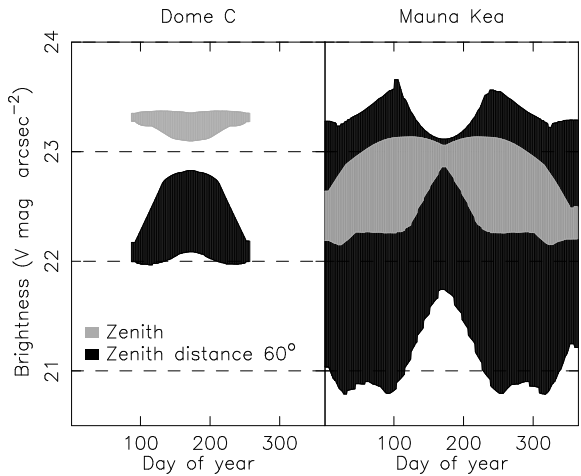}
  \caption{The range of zodiacal contribution to sky brightness (V mag
  arcsec$^{-2}$) at Dome C (left) and Mauna Kea (right), over 1 year,
  for sun elevations less than $-18^\circ$. The grey shading is for
  observations at zenith and the black is for a zenith distance of
  60$^\circ$ at the same azimuthal position as the sun. \label{fig:zodiaczenith}}  
\end{figure}

The zodiacal contribution to sky brightness at Dome C and Mauna Kea
was calculated over one year, using the method described by
\citet{Leinert1998}. All correction factors in the model were set to
unity; thus there is a possible variation in flux of up to
30\%. Calculations were limited to astronomical nighttime. Two sky
positions were looked at: the zenith, and a zenith distance of
60$^\circ$ at the same azimuthal position as the sun. The results are
shown in Figure \ref{fig:zodiaczenith}. 
\placefigure{fig:zodiaczenith}

In the V-band, the zenith brightness
of zodiacal light at Dome C is always \emph{darker} than 23.1
mag arcsec$^{-2}$ because the zodiac is always fairly low on the
horizon. In comparison, at Mauna Kea the zodiacal sky brightness at
zenith is always \emph{brighter} than about 23.1 mag arcsec$^{-2}$, reaching a
maximum contribution of about 22.1 mag arcsec$^{-2}$. At a zenith
distance of 60$^\circ$ and in the azimuthal direction of the sun, the zodiacal
light at Dome C is always darker than about 22 mag arcsec$^{-2}$; at
Mauna Kea it can get as bright as about 20.8 mag arcsec$^{-2}$. This
reduced contribution from zodiacal light is of course not a
characteristic of the site per se, but rather a reflection of the fact
that a different part of the sky passes overhead at Dome C than does
at Mauna Kea. 

\subsection{Airglow}
Airglow, at night, is the chemiluminescence of upper atmosphere
molecules and atoms. This so-called ``nightglow''
includes a quasi-continuum from NO$_2$ and a number of 
discrete emission lines, the strongest by far being from the hydroxyl
radical. A number of visible nightglow emission lines are listed in Table
\ref{tab:airglow}, along with the typical altitude of the emitting atmospheric
layer and typical intensities. 

Nightglow is unpolarised and on a large scale increases
from zenith to the horizon, as described by the van Rhijn function.
On a small spatial scale nightglow emissions are
uneven and blotchy across the sky. Nightglow emissions vary over short and long time
scales. 
\placetable{tab:airglow}
\begin{deluxetable}{llrrl}
  \tablewidth{0 pt}
  \tablecaption{Airglow emissions at zenith in the visible range\label{tab:airglow}}
  \tablehead{\colhead{Source}& \colhead{Wavelength (nm)} & \colhead{Height (km)} &
     \colhead{Typical intensity (R)} & \colhead{Comments}}
    \startdata
    \sidehead{Nightglow from chemical association}
O$_2$    & 260 -- 380   & 90         & 0.5R/\AA  & Herzberg bands\\
OH       & 600 -- 4500  & 85         & 4.5 MR    & Strongest bands in NIR\\
NO$_{2}$ & 500 -- 650   & 90         & 250       & Nightglow continuum\\
\ion{O}{1} & 557.7      & 95         & 250       & \\
Na D     & 589.0, 589.6 & $\sim92$   & 50        & Strong seasonal variation\\
O$_2$   & 761.9         & $\sim80$   & 1000      & Atmospheric bands\\
O$_2$   & 864.5         & $\sim80$   & 1000      & Atmospheric bands\\
\sidehead{Nightglow from ionic reactions}
N       & 519.8, 520.1  & 1    \\
\ion{O}{1}    & 557.7   & 250 -- 300         & 20  & High atmospheric, chiefly in tropics\\
      &       &         & &     during enhancement of OI 630.0,636.4\\
\ion{O}{1}    & 630.0 & 250 -- 300 & 100 & Sporadic enhancements in tropics\\
\ion{O}{1}    & 636.4 & 250 -- 300 & 20  & assoc. with ionospheric disturbances\\
\enddata
\tablerefs{\cite{Roach}, \cite{Leinert1998} and references therein}
\tablecomments{1R=$1/4\pi\times 10^{10}$ photons m$^{-2}$ s$^{-1}$ steradian$^{-1}$}
\end{deluxetable}

Nightglow emissions are mainly from the thin mesospheric layer,
centred at an altitude of 85 -- 90 km. OH
Meinel bands are primarily excited by a reaction between ozone and atomic
hydrogen \citep{texier}. The concentrations of O$_3$ and H depend
on the atomic oxygen mixing ratio \citep{texier} which is
largely controlled by the transport processes in the mesosphere and
lower thermosphere \citep{Garcia1985}. At low and mid-latitudes there
is a semi-annual variation in atomic hydrogen and hence OH nightglow
\citep{texier}. 

Because of the efficient mechanisms that transport
reactants from sunlit locations to the poles, there is no reason to
expect a diminution in the chemiluminescence of species such as OH and
O$_2$ during the long polar night. \citet{Phillips1999} observed a
small reduction in OH emission at South Pole in 1995 
relative to temperate sites, but this is more likely explained by the
known highly variable nature of OH emission. Continuous monitoring of
the Meinel bands of OH in J band with a Michelson interferometer has
been carried out at the Pole since January 1992, and the data are now
publicly available \citep{Sivjee2005}. As expected, these data show
very large hourly and nightly variations, with no diminution in
average intensity as the winter progresses.

\citet{Zaragoza} measured the OH nightglow emissions in a narrow
spectral band near 4.6 $\mu$m for just under one year. Using the
Improved Stratospheric and Mesospheric Sounder on the UARS satellite
they have almost full global coverage (80$^\circ$ N to 80$^\circ$
S). They find at high latitudes a springtime minimum in the OH
emission $\sim 30$\% below the global mean, although this is only for
one year of data. During the Northern hemisphere winter solstice
period their averaged measured intensities of OH nightglow at high
latitudes and mid-latitudes are the same to within the errors. 

Measurements of OH nightglow between 837.5 and 856.0 nm at Davis, 
Antarctica ($68^\circ35'$ S, $77^\circ58'$ E) over 7 years
\citep{Burns2002}, show a barely significant seasonal variation in
emission intensity, although there are large day-to-day variations (Gary Burns
\& John French, 2005, private communication). We therefore expect
that the sky brightness in those bands (650 nm -- 2.2 $\mu$m)
dominated by OH emission will be essentially identical at Dome C to
that at all other observatory sites, including Mauna Kea. 

The strongest nightglow emission in the visible is the 557.7 nm line of
O($^1$S). As discussed in \citet{Shepherd1997}, the
557.7 nm emission is caused by the following reactions:
\begin{eqnarray}
  \text{O}_2+hv & \rightarrow & \text{O}+\text{O}\\
  \text{O}+\text{O}+\text{M} & \rightarrow & \text{O}^*_2\\
  \text{O}^*_2+\text{O} &\rightarrow & \text{O}_2+\text{O}(^1\text{S})\\
  \text{O}(^1\text{S})  &\rightarrow & \text{O}(^3\text{P})+hv\text{ (557.7 nm)}
\end{eqnarray}
where M is usually N$_2$. O$^*_2$ and O($^1$S) have very short
lifetimes, however atomic oxygen above 100 km altitude has a long
lifetime and can be transported globally before recombination.

Using a model of global tides in the thermosphere, ionosphere and
mesosphere, \citet{Yee1997} predict an \emph{increase} of some 50\% in the
O($^1$S) 557.7 nm emissions towards the poles, although more data are
needed to confirm this model. Note, however, that this 
single strong line can be easily filtered from astronomical
measurements. In general, we expect that there will be little
difference in overall airglow emissions between Antarctic and
temperate sites.

\subsection{Integrated starlight, diffuse galactic light and integrated cosmic light}
Telescopes can only identify, as individual objects, those stars that
are brighter than a certain magnitude. The integrated flux from all
stars fainter than this magnitude contributes to the sky
background. The limiting magnitude of a telescope depends on the
\emph{seeing} of the site, the atmospheric extinction and the size of
the telescope. The expected excellent seeing and low atmospheric
extinction at Dome C will ensure that the limiting magnitude of even a
small telescope will be sufficient to reduce the integrated unresolved
starlight to negligible levels. 

Diffuse galactic light is a result of scattering of starlight by
interstellar dust and is brightest in the Milky Way where the
concentrations of stars and dust are highest. The typical zenith value of
diffuse galactic light in the V band is about 23.6 mag arcsec$^{-2}$
\citep{Roach}. At Dome C the galactic
plane continuously circles close to the zenith, leading to a relatively
high contribution from integrated starlight and diffuse galactic light
compared to other sites. At Dome C the galactic latitude of the zenith
ranges between $-10^\circ$ to $-40^\circ$, whereas at Mauna Kea the
range is much larger: $-40^\circ$ to $+85^\circ$ with the galactic plane
close to the horizon for some of the time. 

The integrated cosmic light (redshifted starlight from unresolved
galaxies) contribution, at all sites, is very small in comparison
to all other sources of sky brightness. No firm measurement exists
for its value, however upper and lower limits have been measured and
estimated from models, giving a range of about 25 to 30
mag arcsec$^{-2}$ at 550 nm (Leinert et al. 1998 and references therein).

Dome C will receive a comparatively higher contribution of light from
the galactic plane than lower latitude sites. However this
contribution is small when compared to other sources of sky
brightness and, as with zodiacal light, depends on the sky position
being observed.

\subsection{Light Pollution}
Light pollution from towns and cities can cause a considerable
increase in the brightness of the night sky. Light pollution is mostly
caused by light from vapour lamps (Hg-Na emission lines in blue-visible)
and incandescent lamps (weak continuum)
\citetext{\citealt{Benn1998} and references therein}. \citet{Garstang1989a} 
modelled the increase of the zenith sky brightness caused by light
pollution at various sites. \citet{Garstang1989b} further predicts the
increase in light pollution at various sites over time. For example,
\citet{Garstang1989b} predicts the artificial sky brightness
will increase the total sky brightness by between 0.04 and 0.53 mag
arcsec$^{-2}$ at Mauna Kea by the year 2020. This estimate is based 
on projections of population increase 
and does not take into account changes in light sources and
systems or unforeseen circumstances such as major tourist or
housing developments nearby the sites, as noted in the paper. Light
pollution can significantly increase the brightness of the night sky. For further
discussions of night sky brightness modelling and world maps of the
artificial sky brightness see \citet{Cinzano2001, Cinzano2004} and
references within. 

The closest station to Dome C (Vostok at $78^\circ27'51''$S
$106^\circ51'57''$ E) is 560 km away. The placement of
external lighting at the Dome C station itself will be carefully considered in
relation to astronomical observations. With proper planning there
should continue to be no artificial light pollution at Dome C, while the
light pollution at other sites is likely to increase.

\section{Atmospheric Extinction}\label{sec:atmos}
Atmospheric extinction is caused by the scattering and absorption of
light as it travels through the atmosphere. Extinction in the
atmosphere decreases the amount of light received by an observer and
lowers the limiting magnitude of a telescope. 

As radiation travels from the top of the atmosphere to the
surface of the Earth it may be transmitted directly to the surface with no
attenuation or scattering. Alternatively, it may undergo one
or more of the following processes: single or multiple
scattering, reflection from the
Earth's surface, and absorption by an atmospheric particle or
molecule. Scattering is described by Rayleigh scattering ($2\pi
r\ll\lambda$) from gaseous
molecules, Mie scattering ($2\pi r\sim\lambda$) from aerosol
particles (where $r$ is the radius of the particle and $\lambda$ is
the wavelength of the light interacting with the particle), and a small amount of scattering due to turbulent
cells. Radiation is absorbed 
in the atmosphere by molecules (especially water vapour and ozone),
aerosol particles and liquid or frozen water in clouds \citep{Coulson}. All these
processes of scattering and absorption are strong functions of
altitude and wavelength and their effect, except for Rayleigh scattering, is
highly variable with location and time. 

Scattering has two effects on
astronomical observations: it increases the overall sky
brightness and decreases the flux received from the object being
observed. Atmospheric absorption will also decrease the signal flux. Beer's
law describes the attenuation of light caused by the atmosphere:
\begin{equation}\label{eq:beers_law}
I_{\lambda}=I_{0}\exp(-\tau_{\lambda}^{R}-\tau_{\lambda}^{A}-\tau_{\lambda}^{W}-\tau_{\lambda}^{O}-\tau_{\lambda}^{C})
\end{equation}
where $I_\lambda$ is the flux received by the observer, $I_0$ is the
flux outside the atmosphere and $\tau$ is the wavelength-dependent
optical thickness of the atmosphere; a sum of the Rayleigh $R$,
aerosol $A$, water vapour $W$, ozone $O$ and cloud $C$ optical
thicknesses. 
\subsection{Rayleigh Optical Depth}
Rayleigh optical depth is dependent on the molecular composition of the
atmosphere and is proportional to
$\lambda^{-4}$. The molecular composition varies with the pressure,
temperature and refractive index of the atmosphere above the Earth's
surface. Rayleigh optical depth is essentially constant between sites of
similar elevation. See \citet{Bodhaine1999} and \citet{Tomasi2005} for
further details on Rayleigh optical depth calculations.

\subsection{Aerosol optical depth}
Aerosol optical depth is extremely variable with geography and time
and is dependent on the concentration, size, refractive index and
chemical composition of the aerosol particles. The complex refractive 
index of a particle is $m=n-ik$, where $n$ is the real part (1.45 -- 1.60)
 and $k$ (0.001 -- 0.1)
the imaginary part; the bracketed ranges are typical values for most
``dry'' atmospheric aerosols \citep{Coulson}. High humidity in the air
can cause condensation to occur on aerosol particles, which changes
the index of refraction, size and mean density of the
particle. The vertical profile of the aerosol content of the atmosphere can be 
derived from intensity measurements during twilight as different
levels of the atmosphere are illuminated by sunlight. For the best
sites the aerosol optical depth needs to be as small as possible.

No aerosol measurements have been taken yet at Dome C; however aerosol
measurements at South Pole and Mauna Loa, Hawaii, have been taken
since 1974 (\citet{Bodhaine1995} and references therein). The wavelength-dependent aerosol
scattering coefficient is defined as $\beta_{s}=N_s\sigma_{s}$,
where $N_s$ is the number concentration of aerosol particles and
$\sigma_s$ is the scattering cross section. At the South Pole
$\beta_s$(550 nm) is very low, varying between (1 -- 4)$\times
10^{-7}$ m$^{-1}$. The maximum values are seen in winter and are
associated with long-range mid-tropospheric transport of sea salt from
the coast. Scattering from polar stratospheric clouds may also
contribute to the maximum seen in winter \citep{Collins1993}. In
comparison, at Mauna Loa, $\beta_{s}$(550 nm) varies between (6 --
60)$\times 10^{-7}$ m$^{-1}$, showing maximum values during winter and
spring, associated with dust transport.

\subsection{Absorption optical depths}
The absorption of visible and IR radiation in the atmosphere is dependent
on pressure, temperature and the absorbing gas (largely water vapour, ozone)
and aerosol concentrations; all of
which vary with location, height and time \citep{Zuev}. Atmospheric
absorption is described by the imaginary part $k$ of the refractive index
of the atmospheric particles, such that the absorption coefficient
(m$^{-1}$) is
\begin{equation}
\beta_{ap}=\frac{4\pi k}{\lambda}=\sigma_{a}N_{a}
\end{equation}
where $\sigma_{a}$ is the absorption cross section (m$^2$) and
$N_{a}$ is the concentration of absorbers (m$^{-3}$). At most
wavelengths water vapour is the primary absorber in the atmosphere and
its concentration is highly variable. The optical thicknesses for
water vapour and clouds show complicated and highly variable changes
in magnitude with wavelength and time. The aerosol absorption
coefficient for light at 550 nm at South Pole varies between $2\times
10^{-10}$ and $5\times 10^{-7}$ m$^{-1}$, comparatively lower than at
Mauna Loa (from $1\times10^{-8}$ to $3\times 10^{-7}$
m$^{-1}$)(Bodhaine 1995 and references therein).

To minimise the total optical depth, a site with low water vapour
concentration, few clouds and low aerosol concentration is
required. Dome C fits all these criteria, and may have an even lower
atmospheric aerosol content than South Pole because of its greater
distance from the coast. We therefore expect both the scattering and
absorption by aerosols to be typically an order of magnitude less at
Dome C than at Mauna Kea, and that the overall atmospheric extinction
will be the minimum possible. 

At ultraviolet wavelengths, the reduced atmospheric aerosol content
should lead to improved transmission. However, the infamous ``ozone
hole'' is unlikely to be of significant benefit to astronomers. From
January until August the column density of ozone in the atmosphere
above the South Pole is typically the same as at other sites around
the world. It is only during the spring months that the ozone content
falls to as low as 40\% of its normal value. Unfortunately, these low
ozone values do not occur in the dark winter months. Furthermore,
because the Hartley bands of ozone are heavily saturated, even a
reduction in ozone column density by a factor of three (i.e., somewhat
greater than is actually observed over the South Pole) would shift the
UV cutoff wavelength of the atmosphere by only about 5 nm.

\section{Using a polariser during twilight}\label{sec:pol}
Dome C experiences long periods of twilight, where the solar
depression angle is between zero and $-18^\circ$. \citet{Baldry2001}
have suggested using a polarising filter to reduce the scattered
sunlight contribution to sky brightness during twilight, to achieve
``dark time'' observing. At sites close to the equator this probably
would not be worthwhile because twilight only lasts a few
hours. However, the use of a filter during twilight in Antarctica
could be very beneficial, as noted in that paper. For example, if dark
time conditions at Dome C could be achieved at a solar depression
angle of 15$^\circ$, the available ``dark'' observing time would
increase by 18\%.  

In the next section we discuss the intensity and polarisation of
twilight and in Section \ref{sec:improv} we review the idea of
\citet{Baldry2001}. In Section \ref{sec:SP} we look at 
measurements of the polarisation of \emph{daylight} at South Pole. In
section \ref{sec:spec} we look in some detail at measurements of the total
polarisation of \emph{twilight} in the atmosphere as a function of solar zenith 
angle $Z_\sun$ and wavelength $\lambda$, and explore the benefit that
might be gained by using a polariser on an Antarctic telescope. 

\subsection{Intensity and polarisation of twilight}\label{sec:intens}
During twilight, the total background light $I_B$ reaching the surface
consists of singly scattered sunlight $I_S$, multiply scattered sunlight
$I_M$, and the night sky illuminance $I_N$. 
\begin{equation}\label{eqn:ib}
I_B=I_S+I_M+I_N
\end{equation}
The night sky illuminance includes all the sources of sky brightness
discussed in section \ref{sec:brightness}, except moonlight and
sunlight. Sunlight becomes polarised in the atmosphere through scattering
interactions with permanent atmospheric gases, variable gases and
solid and liquid particles suspended in the atmosphere. The total
degree of polarisation of twilight is given by:  
\begin{eqnarray}\label{eqn:polarisation}
  P&=&\frac{I_S}{I_B}P_S+\frac{I_M}{I_B}P_M
\end{eqnarray}
where $P_S$ and $P_M$ are the degrees of polarisation of singly and multiply
scattered light. The polarised component of the background night sky
flux is assumed to be negligible with respect to that of the scattered
sunlight.  

Singly scattered sunlight is a combination of sunlight scattered from 
molecules and from aerosols. The polarisation of sunlight singly
scattered from molecules can be modelled \citep[see][]{Coulson}; in
this model, the atmosphere is considered to contain only permanent 
gases with no aerosols, clouds, water vapour or ionised particles; the
scattering particles are assumed to be spherical; and the
refractive index very close to unity.
\begin{figure}[h!]
  \plotone{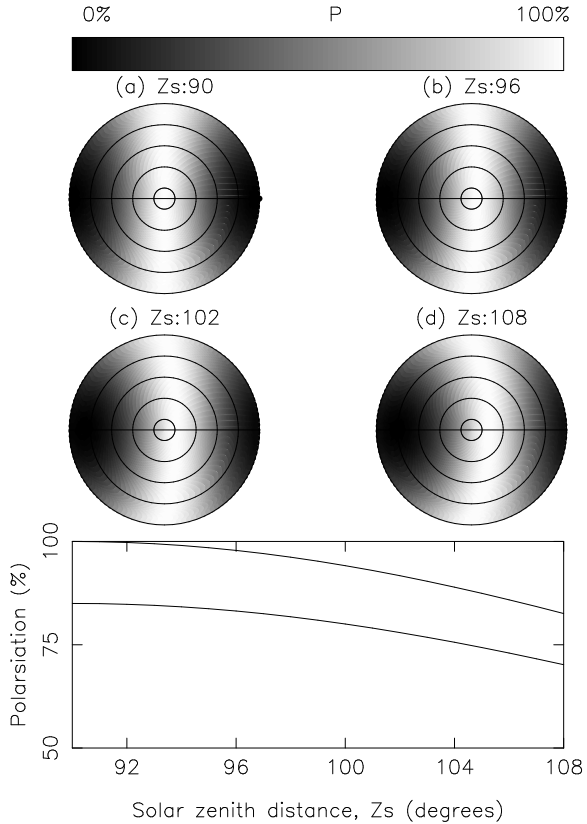}
  \caption{Top: Full sky plots of the degree of polarisation using the
Rayleigh model for sun zenith angles of (a) 90$^\circ$, (b)
96$^\circ$, (c) 102$^\circ$ and (d) 108$^\circ$ (i.e. from sunset to
the end of astronomical twilight) . The circles indicate elevation
angle and are spaced 20$^\circ$ apart. Bottom: Polarisation at zenith
as a function of solar zenith angle for a Rayleigh atmosphere with
assumed maximum polarisation of 100\% and 85\%.\label{fig:rayleigh}}   
\end{figure} 

Figure \ref{fig:rayleigh} shows the results of a
calculation using this model for $Z_\sun=90^\circ$, 96$^\circ$,
102$^\circ$ and 108$^\circ$ (i.e. from sunset 
to the end of astronomical twilight). The light
shading shows that a large section of the singly scattered sunlight
perpendicular to the sun is polarised to a high 
degree, whereas in the direction parallel to the sun the 
polarisation rapidly drops off with zenith angle. In the lower panel
of the figure the polarisation at zenith as a function of increasing
solar zenith angle is shown for $P_{S}^{\text{MAX}}=85$ and
100\%. This indicates that singly scattered light (in an ideal
atmosphere) is highly polarised over a substantial fraction of the
sky during twilight.

The degree of polarisation (although not the pattern) in the real
atmosphere differs from this model because of non-isotropic
molecular scattering, aerosol scattering and reflection from the Earth's
surface.

Multiply scattered sunlight is polarised to a lesser degree than
singly scattered light. Multiple scattering is confined to a thin
atmospheric layer that is much closer to the Earth's surface than the
single scattering layer \citep{Oug1999}. The polarisation of sunlight
can be further reduced by reflection from high level clouds and
cloud-forming particles, e.g. \citet{Pomozi2001}.

The highest polarisation of light occurs when water vapour and aerosol
concentrations are very low and the Rayleigh optical depth is as small
as possible, suggesting that scattered sunlight at Dome C should, in
general, be more highly polarised than at temperate or tropical sites.  
 \placefigure{fig:rayleigh}

\subsection{Potential improvements with a polariser}\label{sec:improv}
 Using an appropriately oriented polariser will reduce the sky background to:
\begin{eqnarray}
  I_B' & = &\tau I_B[1-P(1+2\gamma)]
\end{eqnarray}
where $\tau$ is the transmission of the polariser with unpolarised
light and $\gamma$ is the extinction of the polariser \citep{Baldry2001}.

Assuming photon-noise-limited observations, a perfect polariser and no
change in instrument response with polarisation, a polariser will be
beneficial whenever 
\begin{eqnarray}
P &\gtrsim & (1-\tau)\frac{I_O+I_B}{I_B}
\end{eqnarray}
where $I_O$ is the magnitude of the object. Objects that are very dark
compared to the sky background require $P>52$\% 
for $\tau=0.48$; for a sky background of 20.5 mag arcsec$^{-2}$ and
object of magnitude of 22 mag arcsec$^{-2}$, the polarisation must be
greater than 65\% for the use of a polariser to be beneficial. 

\citet{Baldry2001} assume the polarisation of twilight is always
85\%, independent of solar elevation angle. Although the polarisation
of singly scattered light is probably always above 85\% (for a clear
atmosphere), the total polarisation depends on the relative intensity
of the singly and multiply scattered light to the total sky
intensity, as in Equation \ref{eqn:polarisation}. 

\subsection{Polarisation at the South Pole and effects of ice precipitation}\label{sec:SP}
At this stage no measurements of the polarisation of twilight at Dome C
have been taken. However, \citet{Fitch1983ApOpt} measured the
polarisation of the sky at the South Pole, during summer, under clear
sky conditions, and when ice crystals were evident. They found the
degree of polarisation to be very high, under clear sky conditions, and
close to the results given by a Rayleigh model. The presence of ice
crystals reduced the degree of polarisation and caused a greater
effect at longer wavelengths. We expect that the atmosphere at Dome C
will also closely resemble a Rayleigh atmosphere and therefore, at
small solar depression angles the scattered light will be highly polarised.  

\subsection{Spectral characteristics of polarisation as a function of
solar depression angle} \label{sec:spec}
\begin{figure}[h]
\plotone{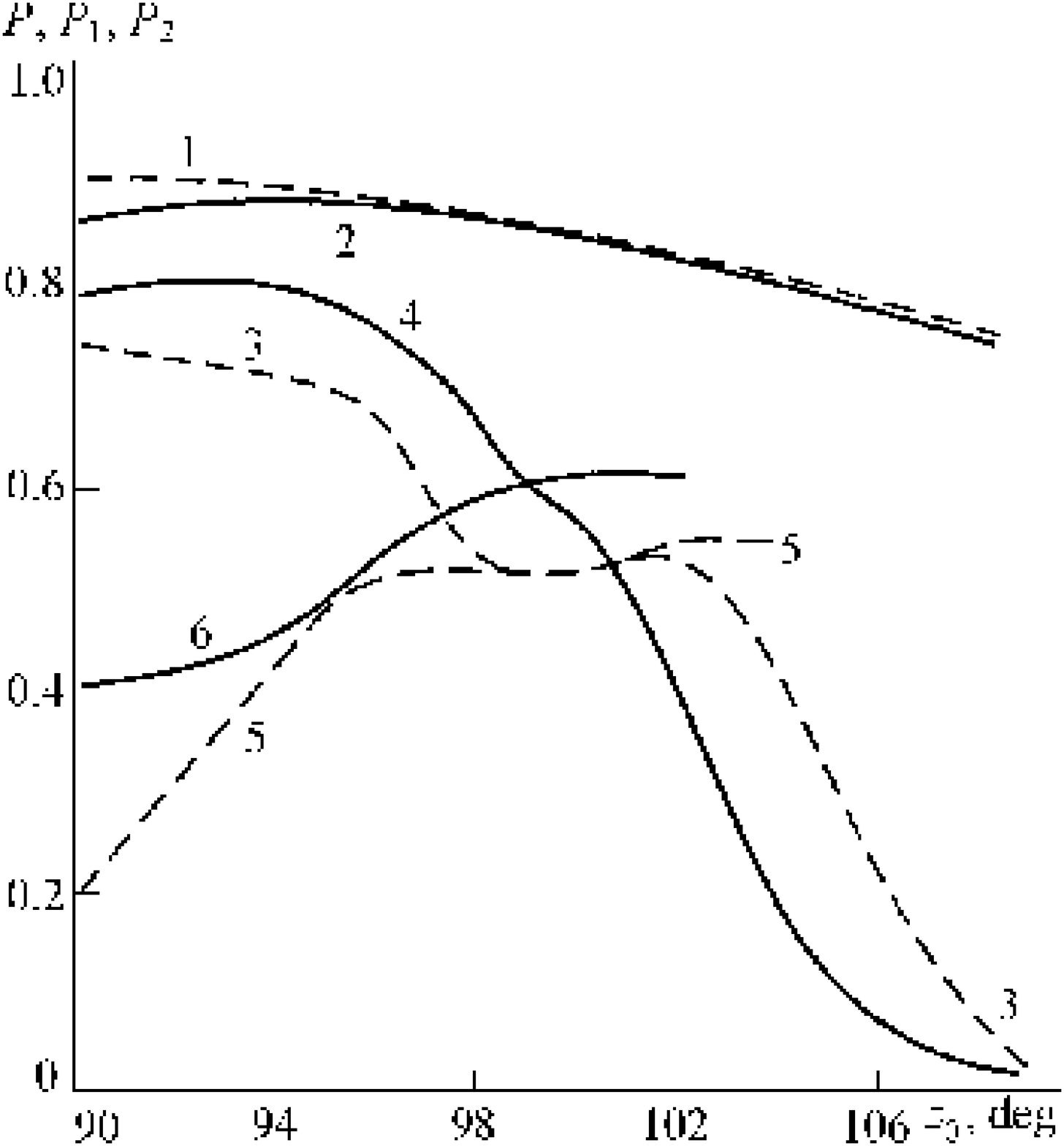}
\caption{Polarisation as a function of solar zenith angle. Curves 1
and 2 are calculated for singly scattered light ($P_S$) in a clear
atmosphere, curves 3 and 4 are averaged measurements of the total
polarisation ($P$), and curves 5 and 6 are the calculated
polarisation for multiply scattered light ($P_M$). The dashed lines
are for polarisation at 480 nm and the solid lines are for 690 nm. (Plot is from \citet{Pavlov1995}).  \label{fig:pavlov}}
\end{figure}

 \placefigure{fig:pavlov}
Over the course of twilight the polarisation and intensity of skylight
changes as different levels of the atmosphere are illuminated.

In this section we look at measurements of the twilight polarisation
and intensity at various high altitude, clean sites. The behaviour of
the total degree of polarisation $P$ as a function of solar zenith
angle $Z_\sun$ can be roughly divided into three regimes; we assess the 
benefit of using a polariser in each regime. $P$ also behaves somewhat
differently for red and blue wavelengths, with a division around
$\lambda=550$ nm. Note that $P(Z_\sun)$ usually shows day to day
variations at each site and there is some overlap in the 3 ranges of
solar zenith angle. This is probably caused by different weather
conditions and vertical aerosol concentrations.

The typical behaviour of the degree of polarisation $P$ as a function
of $Z_\sun$ and $\lambda$ is summarised in Table \ref{tab:pol}, which
is derived from the work of \citet{Bondarenko},
\citet{Coulson1980ApOpt}, \citet{Pavlov1995}, \citet{Ougolnikov2003} and
\citet{Postylyakov2003}. Figure \ref{fig:pavlov}, from
\citet{Pavlov1995}, shows the typical variations of single, multiple
and total polarisation as a function of solar zenith angle during
twilight.  

\placetable{tab:pol}
\begin{deluxetable}{llll}
  \tablewidth{0 pt}
  \tablecaption{Typical behaviour of the degree of polarisation with $Z_\sun$ and $\lambda$.\label{tab:pol}}
  \tablehead{\colhead{} & \colhead{$Z_\sun=90$ -- $96^\circ$} & \colhead{$Z_\sun=96$ -- $102^\circ$} &  \colhead{$Z_\sun=102$ -- $108^\circ$}}
    \startdata 
    $P$ (blue $\lambda$)   & Maximum: $Z_\sun$=92 -- 94$^\circ$	& Steep decrease with increasing	& Steep decrease with \\
                           & Decreases with increasing $Z_\sun$ & $Z_\sun$ then a flat or a minimum  	& increasing $Z_\sun$ then starts  \\
                           & $\lambda$ dependent        	& section between  98 \& 102$^\circ$  	&to  flatten at $Z_\sun\sim104^\circ$\\
    \hline
    $P$ (red $\lambda$)    & Maximum: $Z_\sun=90^\circ$     	& Steady decrease            		&  Steep decrease  \\
                           & Decreases with increasing $Z_\sun$	& Occasional $2^{\text{nd}}$ maximum	&\\
                           & $\lambda$ independent     		& associated with aerosols 		&                   \\
    \hline
    Processes              & $I_S$ dominates over $I_M$ 	& $I_M$ dominates over $I_S$             &$I_S$ dominates over $I_M$ for $Z_\sun >104^\circ$\\
                           & $I_N$ negligible           	& $I_N$ negligible for  $Z_\sun<98^\circ$&$I_N$ dominates over $I_S$ and $I_M$\\
                           &                            	& $P<60$\% for both ranges               &$P$ strongly $\lambda$ dependent  \\
                           &                            	&                                        &Aerosol scattering important \\
                           &                            	&                                        &$P<50$\% for both ranges\\
    \hline
    Polariser              & $Z\sim90^\circ$, $I_O>5.3$       &    $Z\sim98^\circ\phn$, $I_O>14.3$       &    No advantage in using   \\
    advantage              & $Z\sim92^\circ$, $I_O>5.7$        &    $Z\sim100^\circ$, $I_O>18.3$     &   a polariser\\
                           & $Z\sim94^\circ$, $I_O>7.2$        &                                      &   \\
                           & $Z\sim96^\circ$, $I_O>9.9$        &                                      &   \\
\enddata
\tablerefs{\citet{Bondarenko}, \citep{Coulson1980ApOpt}, \citep{Coulson}, \citet{Pavlov1995} }
\tablecomments{The last row shows the maximum object brightness, $I_O$, in
  mag arcsec$^{-2}$ for which a polariser will yield an advantage at
  690 nm, based on the results of \citep{Pavlov1995}.}
\end{deluxetable}

The $Z_\sun=102$ -- 108$^\circ$ regime is where one might have hoped to turn
twilight into dark time.  However, as shown in Table  \ref{tab:pol},
measurements at other sites show 
the polarisation in this range to be less than  about 50\%, and
hence, as discussed in Section \ref{sec:improv}, the use of a
polariser would not be beneficial. The total
polarisation at Dome C at these solar zenith angles is 
likely to be similar to the sites discussed because all measurements
were taken at clean high altitude sites (therefore with small Rayleigh
optical depths) on visibly clear days. We therefore conclude that there is
likely to be no gain in using a polariser, in this regime.

For observations of bright stars (for example, direct imaging of
exoplanets around host stars of typical magnitudes $V=11$) a high sky
background can be tolerated. Using a polariser for this type of observation 
could be advantageous during those times when the sun is up to 6 degrees
below the horizon. 

\section{Conclusions}
Dome C has a comparable number of cloud-free, astronomically dark
hours to a more temperate site such as Mauna Kea. Nevertheless, the
fraction of sky observable at Dome C is considerably lower than at
Mauna Kea. Atmospheric scattering at Dome C should be close to the
lowest anywhere on Earth, reducing the sky brightness 
contributions from sunlight, moonlight and tropospheric
scattering, and reducing the extinction throughout the optical. The
moonlight contribution to sky brightness over the year is
less than at lower latitude sites. Aurorae will rarely
be more than 7$^\circ$ above the horizon and will typically be more
than about 1160 km away; they 
will generally be unobservable. Zodiacal light is darker
at the zenith and 60$^\circ$ from zenith than at equatorial sites 
and will always be darker in V than 23.1 mag arcsec$^{-2}$ at
zenith. Airglow is essentially the same at all sites. The integrated
starlight and diffuse galactic light will be slightly brighter at Dome C
than at other sites because the galactic plane is always close to
zenith. There is no artificial light pollution at Dome C; a condition
that should persist indefinitely. 

In early evening twilight and late morning twilight, some advantage
could be gained through the use of polarising filters. However, as the
sky becomes darker, such filters are of less benefit.

Dome C thus appears to be an attractive site for optical as well
as infrared astronomy. Versatile
facilities such as the proposed two metre telescope PILOT (Pathfinder
for an International Large Optical Telescope \citep{Burton2005})
should therefore be able to achieve their scientific potential at Dome
C across the full observable spectrum. 

\acknowledgments
This research is supported by the Australian Research Council. SLK is supported by an
Australian Postgraduate Award and by an Australian Antarctic
Division top-up scholarship. The authors would like to thank Anna
Moore for helpful discussions and Eric Aristidi for permission to
quote the results of the University of Nice group prior to
publication. We would especially like to thank Gary Burns for very
helpful comments on an earlier draft, and the referee, Ferdinando
Patat, for his useful suggestions which have significantly improved the paper.


\end{document}